\documentclass[iop]{emulateapj}

\usepackage{amsmath}
\usepackage{amssymb}
\usepackage{upgreek} 
\usepackage[colorlinks=true]{hyperref}
\usepackage{braket} 
\usepackage{graphicx}

\let\oldhat\hat
\renewcommand{\vec}[1]{\boldsymbol{#1}} 
\renewcommand{\hat}[1]{\oldhat{\boldsymbol{#1}}} 

\providecommand{\abs}[1]{\lvert#1\rvert} 
\providecommand{\norm}[1]{\lVert#1\rVert} 
\DeclareMathOperator{\erfc}{erfc} 

\slugcomment{to appear in ApJ}

\shorttitle{Model for Hourglass Magnetic Field}
\shortauthors{Ewertowski \& Basu}

\begin{document}


\title{A Mathematical Model for an Hourglass Magnetic Field}


\author{Bartek Ewertowski and Shantanu Basu}
\affil{Department of Physics and Astronomy, Western University, London, Ontario, N6A 3K7, Canada}
\email{basu@uwo.ca}




\begin{abstract}
\noindent
Starting with a mathematical boundary value problem for the magnetic vector potential
in an axisymmetric cylindrical coordinate system, we derive a general solution for any 
arbitrary current distribution using the method of Green's functions. We use this to 
derive an analytic form for an hourglass magnetic field pattern created by electrical
currents that are concentrated near (but not confined within) the equatorial plane of 
a cylindrical coordinate system. Our solution is not characterized by a cusp at the 
equatorial plane, as in previous solutions based on a current sheet. The pattern 
we derive provides a very good fit to hourglass magnetic field patterns emerging from 
three-dimensional numerical simulations of core formation, and can in principle be used 
for source-fitting of observed magnetic hourglass patterns. 
\end{abstract}



\section{Introduction}
An hourglass-shaped magnetic field is a natural outcome of gravitationally-driven 
contraction of a plasma that is embedded in a large-scale 
magnetic field. Self-inductance and high conductivity of the plasma leads to a flux-freezing
effect that drags the magnetic field lines inward in the dense contracted regions. Hourglass
patterns are evident in theoretical calculations of magnetohydrostatic equilibria  
\citep{mou76} and in dynamical calculations of collapsing magnetized molecular cloud cores
\citep[e.g.,][]{kudoh_three-dimensional_2008,kudoh_formation_2011}. The first unambiguous observed hourglass pattern in a star-forming
region was in NGC 1333 IRAS 4A \citep{girart_magnetic_2006}. This was measured by submillimeter dust emission
that is polarized perpendicular to the direction of the ambient magnetic field. Subsequent dust 
polarization measurements also show an hourglass pattern in the high-mass star-forming
region G31.41+0.31 \citep{girart_magnetic_2009}.

These measurements confirm the general scenario of gravitational collapse in regions with
large-scale magnetic fields, and that the turbulent energy in the plasma does not dominate
the magnetic energy. The ambient strength of the magnetic field is however not measurable
through the polarization signal, due to uncertainties in grain properties that lead to 
alignment and emission strength. The {\it shape} of the hourglass pattern does however 
hold the potential to constrain the magnetic field strength (see \citet{basu_nonlinear_2009} for a 
discussion of this point based on numerical models). The degree of pinching of the hourglass
contains information about the ratio of central to background magnetic field strength.
If this is combined with information about the ratio of densities in these regions, one can 
estimate the relative degree of flux-freezing in the contraction of the star-forming
region. By this indirect means, one can determine the degree of flux-freezing and hence the
degree of ambipolar diffusion (neutral-ion drift) during the contraction. 
There are essentially two limits of core formation in a magnetized environment: (1) contraction
due to neutral-ion drift (with low amounts of inward field-line dragging) 
in a magnetically-dominated environment, and (2) gravitationally-dominated contraction
with significant inward field-line dragging. The former occurs on the ambipolar diffusion
time scale and the latter on the (shorter) dynamical time scale. These are well illustrated
through linear analyses of gravitational instability in magnetized media \citep{cio06,mou11,bai12}. There is a continuum of possibilities between the above
two limits, with increasing amounts of pinching of the hourglass while
progressing from the first to second limit. 

In this paper, we find an analytic form for an hourglass magnetic field pattern, extending
previous studies \citep{mestel_disc-like_1985,barker_disc-like_1990} that have calculated patterns based on a current sheet in the 
equatorial plane of a cylindrical coordinate system. Unlike those solutions, the pattern 
derived here does not
have a cusp at the midplane, and is more generally able to fit results of three-dimensional
numerical simulations. The intent of this work is to provide analytic formulae that can be used relatively easily for observational 
source-fitting and characterizing properties of the magnetic field, 
e.g., the ratio of central to background field strength. 
Our model is formally suited to fit the hourglass magnetic field pattern of a prestellar core, however it could also be applied to protostellar cores in some 
cases, as discussed in Section~\ref{discussion}.

A theoretical hourglass pattern can also be used
to generate synthetic polarization emission maps, as done recently by \citet{kat12} using 
numerical models of hourglass fields. Their numerical models also include a toroidal 
magnetic field that is generated by rotation, and they point out the potential 
of source-fitting in order to gain information about magnetic field strength, rotation
rate, and evolutionary state.

\section{Model}

Following \citet{mestel_disc-like_1985}, we assume that the magnetic field threading a flattened star-forming core (hereafter referred to as ``the disk") can be considered as a local distortion of an initially uniform background field $\vec{B}_0$. We introduce cylindrical coordinates $(r, \varphi, z)$, with the origin at the center of the protostellar disk and the $z$-axis parallel to the background field. Thus, the background field is 
\begin{equation}
\vec{B}_0 \equiv B_0 \, \hat{z} \,,
\label{eq:B0}
\end{equation}
where $B_0$ is a constant. We express the net magnetic field $\vec{B}$ as 
\begin{equation}
\vec{B} \equiv \vec{B}_d + \vec{B}_0 \,,
\label{eq:B_total}
\end{equation}
where $\vec{B}_d$ represents the local magnetic field generated by the disk currents.

We require that the magnetic field through the disk is axisymmetric and thus define
\begin{equation}
\frac{\partial{}}{\partial{\varphi}} \equiv 0 \,.
\label{eq:cylsym}
\end{equation}
In order to produce a simple hourglass morphology without any azimuthal twisting in the field lines, we also require that $\vec{B}$ is purely poloidal, that is 
\begin{align}
\vec{B} \equiv B_r \, \hat{r} + B_z \, \hat{z} \,.
\label{eq:B_P}
\end{align}
Since $\vec{B}$, and thus $\vec{B}_d$, must be solenoidal, we write
\begin{equation}
\vec{B}_d=\nabla \times \vec{A} \,,
\end{equation}
where $\vec{A}$ is a purely toroidal Coulomb gauge vector potential, given by
\begin{align}
\vec{A} \equiv A \, \hat{\varphi} \equiv \frac{\Phi}{r} \, \hat{\varphi} \,.
\label{eq:A}
\end{align}
The function $\Phi(r,z)$ is the poloidal magnetic flux (divided by a factor of $2 \pi$) due to $\vec{B}_d$, and can be expressed as
\begin{equation}
\Phi \equiv \int_0^r \vec{B}_d \cdot \hat{z} \, r^\prime dr^\prime \,.          \label{eq:Phi}
\end{equation}
Hereafter we will refer to $\Phi$ simply as ``the flux." For a derivation of Eq. (\ref{eq:A}), see \citet{kulsrud_plasma_2004}.

In general, using the pre-Maxwell form of Amp\`ere's law (expressed in Gaussian cgs units), the Coulomb gauge vector potential $\vec{A}$ produced by a volume current density $\vec{j}$ satisfies
\begin{align}
- \nabla^2 \vec{A} =& ~\frac{4\pi}{c}\vec{j} \,,	\label{eq:ampere} \\
\nabla \cdot \vec{A}=& ~0 \,.		\label{eq:A_solenoid}		
\end{align}
Given our expression for $\vec{A}$ in Eq. (\ref{eq:A}), it follows that the volume current density in the disk must be purely toroidal as well, and we write
\begin{align}
\vec{j} \equiv j \, \hat{\varphi} \, .
\end{align} 
Thus, Eq. (\ref{eq:ampere}) simplifies to
\begin{align}
	-\nabla^2 A + \frac{A}{r^2} = \frac{4\pi}{c} j \,.
	\label{eq:PDE}
\end{align}

We treat $r$ and $z$ as ordinary Cartesian coordinates, and introduce the operator $\tilde{\nabla}$ with respect to these variables:
\begin{align}
	\tilde{\nabla} \equiv \frac{\partial{}}{\partial{r}} \hat{r}  + \frac{\partial{}}{\partial{z}} \hat{z} \,.
	\label{eq:nabla_2D}
\end{align}
Multiplying Eq. (\ref{eq:PDE}) by a factor of $r$, we express the PDE in the form
\begin{align}
	L\,A \equiv -\tilde{\nabla} \cdot \left( r \, \tilde{\nabla} A \right) +\frac{A}{r} = \frac{4\pi}{c} r \, j \,,
	\label{eq:PDE_SL}
\end{align}
where $L$ is a self-adjoint elliptic operator. This greatly simplifies the task of finding the Green's function for this problem.

We impose the boundary conditions that $\vec{B}$ must relax to the background field $\vec{B}_0$ as $z\rightarrow \pm \infty$, and likewise for $r \geq R$, where $R$ is the disk radius. Thus, we require that the locally generated flux satisfies
\begin{align}
\Phi(0,z)=& \,0 \,, \\
\Phi(R,z)=& \,0  \,, \\
\lim_{z\rightarrow \pm \infty} \Phi(r,z)=& \,0 \,.
\end{align}
Using Eq. (\ref{eq:A}), we translate the preceding conditions on $\Phi$ into equivalent conditions on $A$:
\begin{align}
	A(0,z)& ~~~\mathrm{bounded},			\label{eq:BC1}\\
	A(R,z)&=\, 0 \,,							\label{eq:BC2}\\
	\lim_{z\rightarrow \pm \infty} A(r,z)&= \, 0 \,. 	\label{eq:BC3} 
\end{align}

Given these conditions, we solve Eq. (\ref{eq:PDE_SL}) for $A$ on the domain $D= \left\{ (r,z) \in \mathbb{R}^2 \,| ~r\in(0,R), ~z\in(-\infty, \infty) \right\}$. We do so by finding the Green's function $G(\xi,\eta,r,z)$ for the problem. Thus, we first solve the associated PDE 
\begin{align}
L \,G(\xi,\eta,r,z) = \delta(r-\xi) \delta(z-\eta)
\label{eq:PDE_Green}
\end{align}
for a point source located at $(r,z)=(\xi,\eta)\in D$ with identical boundary conditions (\ref{eq:BC1}) through (\ref{eq:BC3}) imposed on $G(\xi,\eta,r,z)$. Since the coefficients in the operator $L$ and the boundary conditions have no explicit dependence on $z$, we write Eq. (\ref{eq:PDE_Green}) in the form
\begin{align}
	L \,G = -r\frac{\partial^2 G}{\partial z^2} + L_r \,G=\delta(r-\xi) \delta(z-\eta) \,,
	\label{eq:PDE_Green2}
\end{align}
where the self-adjoint operator $L_r$ is defined by 
\begin{align}
L_r G \equiv -\frac{\partial}{\partial r}\left( r \frac{\partial G}{\partial r} \right) +\frac{G}{r} \,.
\label{eq:Lr}
\end{align}

We express the Green's function in terms of the eigenfunction expansion
\begin{align}
	G(\xi,\eta,r,z) = \sum_{m=1}^\infty C_m(\xi,\eta,z)\psi_m(r) \,.
	\label{eq:Green_expansion}
\end{align}
The eigenfunctions $\psi_m(r)$ corresponding to eigenvalues $\lambda_m$ are obtained from the following Sturm-Liouville eigenvalue problem associated with the operator $L_r$:
\begin{align}
	L_r \psi_m=\lambda_m r \psi_m \,,
	\label{eq:Lr_eigenvalue}
\end{align}
for $r\in(0,R)$ with the homogeneous boundary conditions
\begin{align}
	&\psi_m(0) ~~~\mathrm{bounded} \,,		\label{eq:Lr_HBC1}\\
	&\psi_m(R)=0 \,.							\label{eq:Lr_HBC2}
\end{align}
This problem amounts to solving Bessel's differential equation with the prescribed boundary conditions. The normalized eigenfunctions are thus well known \citep{zauderer_partial_2006}, and are given by
\begin{align}
	\psi_m(r)=\frac{\sqrt{2}}{R}\frac{ J_1(\sqrt{\lambda_m} r) }{ \abs{J_2(\sqrt{\lambda_m} R)}} \,,
	\label{eq:Bessel_eigen}
\end{align}
where we use the notation $J_\alpha$ to denote Bessel functions of the first kind of order $\alpha$. The corresponding eigenvalues are
\begin{equation}
\label{eq:Bessel_eigenval}
\lambda_m = \left( \frac{ a_{m,1} }{R}  \right)^2	~~~~~~m\in \mathbb{N} \,,
\end{equation}

where $a_{m,1}$ is the $m^{th}$ positive root of $J_1(x)$ and $a_{m,1}<a_{m+1,1}$. The eigenfunctions are orthonormal with respect to the $r$-weighted inner product. Explicitly, 
\begin{align}
	(\psi_m, \psi_n) \equiv \int_0^R\psi_m(r) \psi_n(r) \,r\,dr =\delta_{mn} \,,
	\label{eq:Bessel_orthonormal}
\end{align}
where $\delta_{mn}$ denotes the Kronecker delta. 

Having expressed the Green's function in terms of the eigenfunction expansion in Eq. (\ref{eq:Green_expansion}), we take advantage of the orthonormality condition by multiplying Eq. (\ref{eq:PDE_Green2}) by $\psi_n(r)$ and integrating with respect to $r$ over the interval $(0,R)$. This yields an ordinary differential equation for the expansion coefficients $C_n(\xi,\eta,z)$:
\begin{equation}
	\label{eq:PDE_Green_7}
	-\frac{d^2 C_n(\xi,\eta,z)}{dz^2}  + \lambda_n C_n(\xi,\eta,z)=\delta(z-\eta) \psi_n(\xi) \,.
\end{equation}
This in turn implies that the coefficients can be expressed in the separable form
\begin{equation}
	C_n(\xi,\eta,z)=K_n(\eta,z)\psi_n(\xi) \,,
	\label{eq:exp_coeff}
\end{equation}
where $K_n(\eta,z)$ is a Green's function determined from the problem
\begin{equation}
	-\frac{d^2 K_n(\eta,z)}{dz^2}  + \lambda_n K_n(\eta,z)=\delta(z-\eta) \,. 
	\label{eq:ODE_Green}
\end{equation}
Using the standard Fourier transform method \citep{zauderer_partial_2006}, we find the solution to be 
\begin{equation}
	K_n(\eta,z) = \frac{e^{-\sqrt{\lambda_n}\, \abs{z-\eta}  }}{ 2 \sqrt{ \lambda_n } }   \,. 	\label{eq:ODE_Green_Solution}
\end{equation}

We now combine Eq. (\ref{eq:Green_expansion}), (\ref{eq:Bessel_eigen}), (\ref{eq:exp_coeff}), and (\ref{eq:ODE_Green_Solution}) to get an explicit expression for the Green's function:
\begin{equation}
	G(\xi,\eta,r,z) = \sum_{m=1}^\infty \frac{ J_1(\sqrt{\lambda_m} r) \, J_1(\sqrt{\lambda_m} \xi) \,e^{ -\sqrt{\lambda_m}\, \abs{z-\eta}  } }{ R^2 \sqrt{ \lambda_m} \, \abs{ J_2(\sqrt{\lambda_m} R) }^2}   \,.		
	\label{eq:GREEN_FUNCTION}
\end{equation}
In terms of the Green's function, the general solution to the PDE (\ref{eq:PDE_SL}) with the associated boundary conditions  (\ref{eq:BC1}) through (\ref{eq:BC3}) on the domain $D$ is
\begin{equation}
A(r,z) =\frac{4\pi}{c} \int \limits_{-\infty}^{\infty} \int \limits_0^R G(\xi, \eta,r,z) \, j(\xi, \eta) \, \xi \, d\xi \, d\eta \,.
\label{eq:general_solution}
\end{equation}
For a derivation of the general form of Eq. (\ref{eq:general_solution}), see \citet{zauderer_partial_2006}.

We assume that the current density has the separable form
\begin{equation}
\label{eq:thin_disk_current}
j(r,z) \equiv f(r) g(z) \,.
\end{equation}

In this paper, we assume that the $z$-dependence of the current can be modelled by the Gaussian function
\begin{equation}
g(z) \equiv e^{-z^2/h^2} \, ,
\end{equation}
where $h$ is a free parameter. This choice results in a model with smooth field lines. In contrast, setting $g(z) \equiv \delta(z)$ would yield a thin disk model related to those presented by \citet{mestel_disc-like_1985} and \citet{barker_disc-like_1990}, which exhibit cusps in the field lines at the midplane.

We insert our expressions in Eq. (\ref{eq:GREEN_FUNCTION}) and (\ref{eq:thin_disk_current}) into Eq. (\ref{eq:general_solution}), which we can then express as
\begin{align}
A = \frac{2}{h \sqrt{\pi}}\sum_{m=1}^\infty & k_m e^{ -h^2 \lambda_m/4 } \, J_1(\sqrt{\lambda_m} \, r) \int \limits_{-\infty}^{\infty} g(\eta) \, e^{ -\sqrt{\lambda_m} \abs{z-\eta} } d\eta \,,
\label{eq:A_intermediate}
\end{align}
where we have defined the coefficients
\begin{align}
	k_m \equiv \frac{2 h \pi^{3/2} e^{ h^2 \lambda_m/4 } }{ c R^2 \sqrt{ \lambda_m } \left[ J_2(\sqrt{\lambda_m} R) \right]^2} \int_0^R   f(\xi) \, J_1(\sqrt{\lambda_m} \xi) \, \xi \,  d\xi \,.
	\label{eq:disk_coeff}
\end{align} 
The extra exponential and constants which appear in Eq. (\ref{eq:A_intermediate}) cancel with their counterparts in the coefficients $k_m$. We introduced these factors in order to simplify the form of the final solution.

The integral with respect to $\eta$ in Eq. (\ref{eq:A_intermediate}) is generally much easier to compute by expressing it as
\begin{equation}
\int \limits_{-\infty}^{\infty} g(\eta) \, e^{ -\sqrt{\lambda_m} \abs{z-\eta} } d\eta = \mathcal{L} \left\{ g(z-u) + g(z+u) \right\} \,,
\end{equation}
where $\mathcal{L}$ denotes the unilateral Laplace transform operator from the $u$ domain to the $\sqrt{\lambda_m}$ domain, defined by
\begin{equation}
\mathcal{L} \theta \equiv \int \limits_0^{\infty} \theta(u) \, e^{ -u \sqrt{\lambda_m} } du \,.
\end{equation}
%
Inserting our choice of $g(z)$ into Eq. (\ref{eq:A_intermediate}) and integrating leads to the final solution
\begin{align}
\label{eq:A_series2}
\begin{split}
	A(r,z) =\sum_{m=1}^\infty & k_m J_1(\sqrt{\lambda_m} \, r) \Bigg[ \erfc \left( \frac{\sqrt{\lambda_m} \, h}{2} +\frac{z}{h} \right) e^{ \sqrt{\lambda_m} \, z } \\
&+ \erfc \left( \frac{\sqrt{\lambda_m} \, h}{2} -\frac{z}{h} \right) e^{ -\sqrt{\lambda_m} \, z } \Bigg] \,.
\end{split}
\end{align}

The magnetic field components, including the local and background contribution, are given by the expressions
\begin{align}
	B_z &= \frac{1}{r}\frac{\partial{(A \, r)} }{\partial{r}} \, , \label{eq:Bz} \\
	B_r &= -\frac{\partial{A} }{\partial{z}} + B_0 \, , \label{eq:Br}
\end{align}
which yield the explicit solutions
\begin{align}
\label{eq:Br2}
\begin{split}
	B_r =\sum_{m=1}^\infty & k_m \sqrt{\lambda_m} J_1(\sqrt{\lambda_m} \, r) \Bigg[ \erfc \left( \frac{\sqrt{\lambda_m} \, h}{2} -\frac{z}{h} \right) e^{ -\sqrt{\lambda_m} \, z }  \\
&-\erfc \left( \frac{\sqrt{\lambda_m} \, h}{2} +\frac{z}{h} \right) e^{ \sqrt{\lambda_m} \, z }  \Bigg] \,,
\end{split}
\end{align}
\begin{align}
\label{eq:Bz2}
\begin{split}
	B_z =\sum_{m=1}^\infty & k_m \sqrt{\lambda_m} J_0(\sqrt{\lambda_m} \, r) \Bigg[ \erfc \left( \frac{\sqrt{\lambda_m} \, h}{2} +\frac{z}{h} \right) e^{ \sqrt{\lambda_m} \, z } \\
&+ \erfc \left( \frac{\sqrt{\lambda_m} \, h}{2} -\frac{z}{h} \right) e^{ -\sqrt{\lambda_m} \, z } \Bigg] + B_0 \,.
\end{split}
\end{align}
These expressions are valid for $(r,z) \in D$. For $r \geq R$, the magnetic field is equal to the background value $\vec{B}_0$. 

Rather than assuming a particular model for $f(r)$ and computing the integral in Eq. (\ref{eq:disk_coeff}) directly, we treat the coefficients $k_m$ as unknown fitting parameters. This approach requires that the series is truncated. Fitting the truncated series corresponds to fitting longer wavelength variations in the data, since the eigenvalues $\lambda_m$ are related to the inverse wavelength of the corresponding eigenfunctions $\psi_m(r)$, and smaller eigenvalues are indexed by smaller $m$-values.

\section{Application}

\begin{figure}
\centering
	\includegraphics[width=\columnwidth]{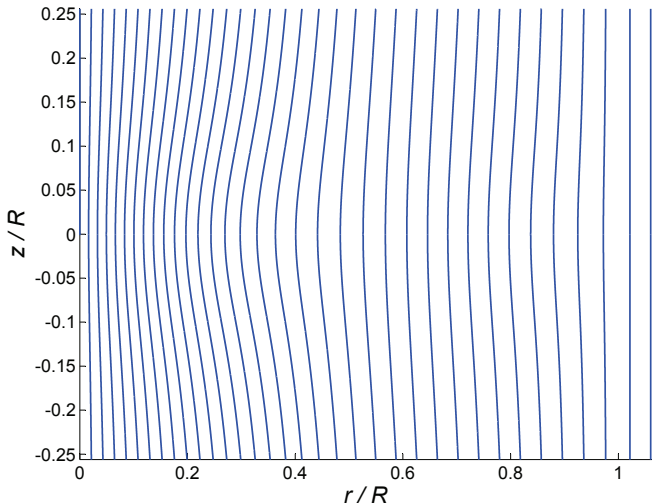} 
	\caption{Plot of the model magnetic field lines. The model parameters are $k_1=0.9549$, $k_2=0.4608$, $k_3=0.6320$, $R=3.8918$, $h=0.3257$, and $B_0=3.3118$, taken from the best fit presented in Fig. \ref{fig:V0_fit}. The values of $R$ and $h$ are measured in the length unit adopted by \citet{kudoh_formation_2011}. The line spacing is inversely proportional to the field strength.}
	\label{fig:field_plot}
\end{figure}
%

\begin{figure}
\centering
	\includegraphics[width=\columnwidth]{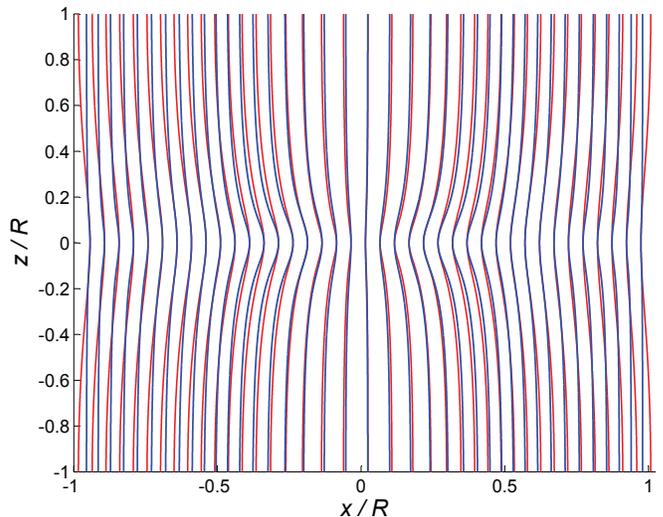} 
	\caption{Plot of the magnetic field lines for the V0 simulation (red lines) and the best fit model (blue lines) in the plane $y=0$. The optimized fitting parameters are $k_1=0.9549$, $k_2=0.4608$, $k_3=0.6320$, $R=3.8918$, $h=0.3257$, and $B_0=3.3118$. The spacing of the lines is determined by the simulation grid and not the field strength.}
	\label{fig:V0_fit}
\end{figure}

\subsection{Sample fit to simulation}
The fitting parameters involved in our model are $R$, $h$, $B_0$ and the coefficients $k_m$. We found that taking only the first three terms in the series for $B_r$ and $B_z$ gave good results. 

To test our model, we fitted magnetic field data from the simulation labeled V0 that is presented by \citet{kudoh_formation_2011}. The data was produced by solving the three-dimensional MHD equations, including ambipolar diffusion and self-gravity. For a complete discussion, see \citet{kudoh_three-dimensional_2008}. The hourglass magnetic field in the V0 simulation exhibits remarkable cylindrical symmetry about the $z$-axis, and has a negligible azimuthal component, making it an ideal data set for our model. 

We took a slice of the data in $y=0$ plane and fitted our model in the least squares sense. That is, we minimized the sum
\begin{equation}
\sum_i \, \norm{\vec{B}(x_i \, , \, z_i) - \vec{B}_{V0}(x_i \, , \, z_i)}^2 \,,
\end{equation}
where $\vec{B}_{V0}(x_i \, , \, z_i)$ denotes the $x$- and $z$-components of the simulation magnetic field at the point $(x_i \, , \, z_i)$ in the plane $y=0$, and $\vec{B}(x_i \, , \, z_i)$ denotes the model magnetic field. We used MATLAB's sequential quadratic programming algorithm to perform the optimization.

The field lines generated using the resulting fitting parameters are plotted in Fig. \ref{fig:field_plot}. The model shows very good agreement with the simulation field line morphology, as illustrated in Fig. \ref{fig:V0_fit}.

\subsection{Fitting polarization data}
Fitting polarization data requires a different approach. Polarization maps, such as the ones published by \citet{girart_magnetic_2006, girart_magnetic_2009}, only convey the morphology of the field lines, but not the field strength. Thus, fitting can be performed by minimizing the angle between the model magnetic field vectors and the vectors obtained from the polarization data. 

The shape of the field lines in our model depends only on the ratio $\vec{B}/B_0$, regardless of the specific value of the background field strength $B_0$. Thus, when fitting polarization data, $B_0$ should be measured independently rather than being determined as a fitting parameter. We should instead fit $\vec{B}/B_0$ to obtain the parameters $R$, $h$, and $k_m/B_0$. Then, provided an estimate for the background field strength, we can calculate the actual $\vec{B}$-values in the region of interest.

Our fitting function can be compared directly to a plane-of-sky polarization map in the manner described above. However, a more sophisticated approach
would be to use a model for thermal dust emission and polarization \citep[e.g.,][]{pad12} to combine our axisymmetric three-dimensional fitting function with a 
model of dust properties in the region. This would lead to a synthetic 
plane-of-sky polarization map that can be compared with the observed 
polarization map.

\begin{figure}
\centering
	\includegraphics[width=\columnwidth]{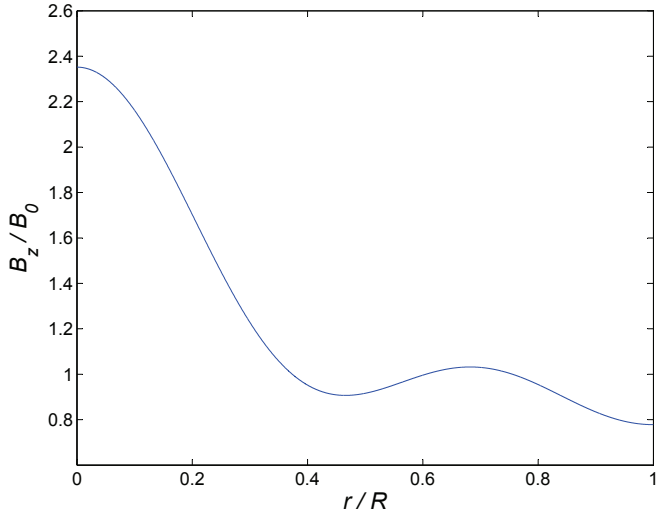} 
	\caption{Plot of the $z$-component of the model magnetic field at $z=0$. The fitting parameters are $k_1=0.9549$, $k_2=0.4608$, $k_3=0.6320$, $R=3.8918$, $h=0.3257$, and $B_0=3.3118$.}
	\label{fig:Bz_plot}
\end{figure}
%

\begin{figure}
\centering
	\includegraphics[width=\columnwidth]{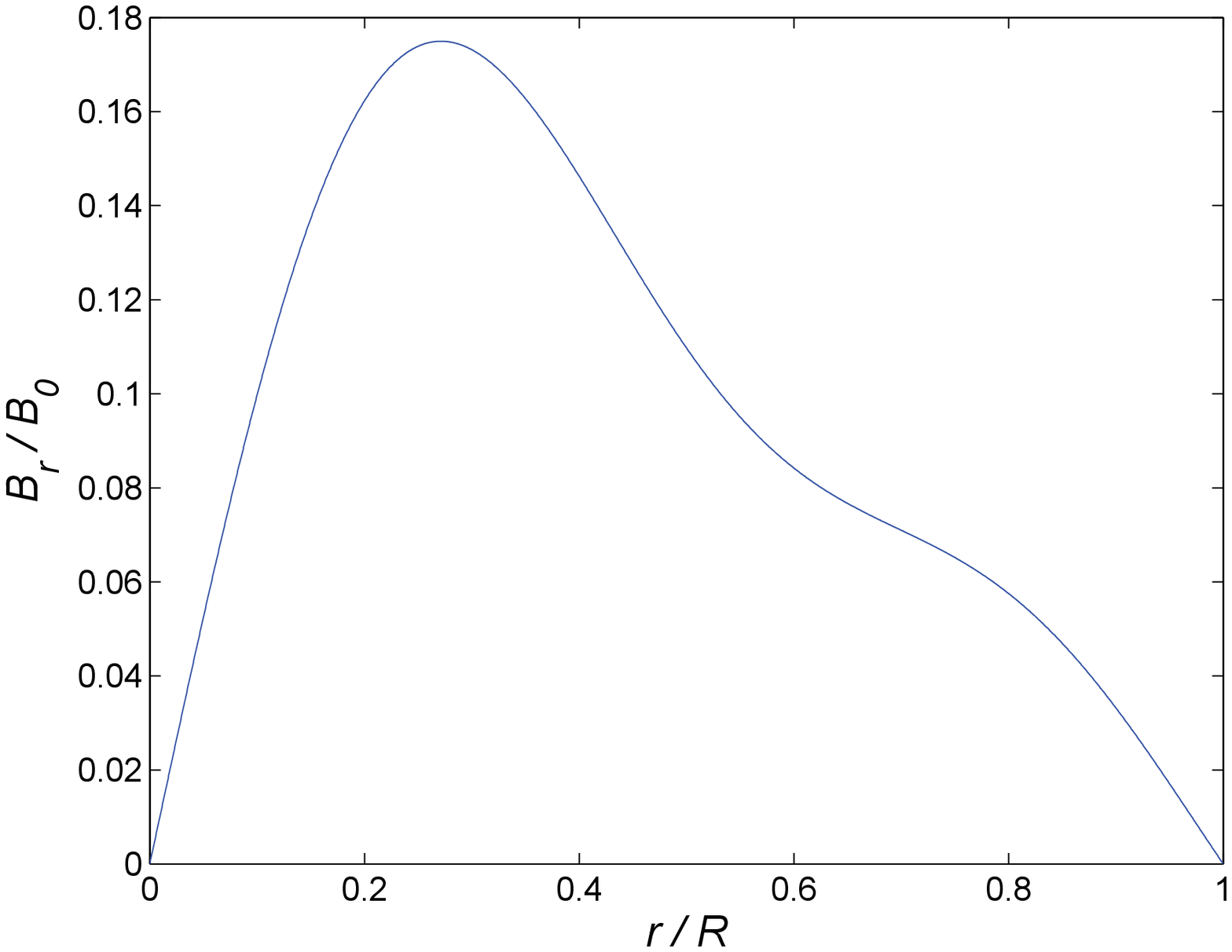} 
	\caption{Plot of the $r$-component of the model magnetic field at $z=0.26\,R$. The fitting parameters are $k_1=0.9549$, $k_2=0.4608$, $k_3=0.6320$, $R=3.8918$, $h=0.3257$, and $B_0=3.3118$.}
	\label{fig:Br_plot}
\end{figure}

\section{Conclusions and Discussion}
\label{discussion}

Our model can be a useful tool for analyzing observational data. Provided that an estimate of the background field is available, the model can be used to predict the central field strength in a cloud. The ratio of central to background magnetic field strengths can also be compared with the ratio of central to background column densities in order to determine the degree of flux freezing.  


Using the Green's function in Eq. (\ref{eq:GREEN_FUNCTION}), it is easy to modify our model by using different assumptions about the current in the disk. For instance, it would certainly be possible to reduce the number of fitting parameters by picking a particular model for the $r$-dependent portion of the current, $f(r)$, and evaluating the integral in Eq. (\ref{eq:disk_coeff}) explicitly. Since the disk has a finite size, $f(r)$ should be chosen so that the function goes smoothly to zero at $r=R$.

Although our model certainly provides an excellent fit to the simulation data of \citet{kudoh_formation_2011}, there is one drawback that should be discussed. As shown in Fig. \ref{fig:Bz_plot}, when the truncated series for $B_z$ is fitted to a data set the resulting function generally has a small discontinuity across $r=R$, the size of which decreases as $z \rightarrow \pm \infty$. This discontinuity occurs because $B_z$ is given by Eq. (\ref{eq:Bz}) and we assume that the disk has a finite size $R$. It is important to note that the truncated series for $B_r$ is always continuous across $r=R$, as illustrated in Fig. \ref{fig:Br_plot}. Consequently, the discontinuity is in no way apparent when plotting the field lines since they are perfectly smooth for all $r$-values. For example, see Fig. \ref{fig:field_plot}.

It is possible to minimize the size of the discontinuity by introducing the optimization constraint that $B_z$ is continuous at $(r,z)=(R,0)$. However, we found that this additional constraint reduced the overall quality of the fit when using three terms in the series. Taking several more terms in the series would likely give better results.

For source-fitting purposes, our Gaussian-disk model has two important 
advantages over the related field line model of \citet{barker_disc-like_1990},
even though both models would yield similar field line patterns at large
distances from the midplane. That model is infinitesimally
thin in the vertical direction, leading to a cusp at the midplane of the 
hourglass. Any real star-forming core will have a finite vertical 
thickness, and our fitting parameter $h$ can be identified with this 
physical thickness. The \citet{barker_disc-like_1990} model also has an 
infinite radial extent. Had we also assumed an infinite radial extent, then the
series over discrete wavelengths in Eq. (\ref{eq:A_series2}) would be 
replaced by an integral over a continuous spectrum. The infinite 
thin-disk model published by \citet{barker_disc-like_1990}, for example, 
requires numerical integration and iteration in order to compute the 
magnetic field. Our truncated series for the magnetic field, on the 
other hand, is much simpler to evaluate and fit to data using basic 
nonlinear least-squares techniques. The fitting parameter $R$ can also 
be identified with the physical radial extent of the star-forming core.

Our model is formally designed to fit the hourglass pattern of a prestellar
core, since the profile of $B_z$ is flat (not cusped) at the center, 
$B_r$ goes to zero at $r=0$, and there is no toroidal field $B_{\phi}$. 
Once a central protostar has formed, the 
protostellar core will have cusped power-laws in $B_z$ and $B_r$ at small 
radii, as well as a significant ratio of $B_{\phi}$ to $B_z$
\citep[see][]{mac07,dap12}.
All of this occurs within the radius of the expansion wave that emanates
from the center after the protostar forms. 
Our model could be applied to a protostellar core if 
the region of the expansion wave is small and/or unresolved. 
Furthermore, depolarization of the inner high density regions may
occur due to a decrease of grain alignment efficiency \citep[see][]{pad12}, 
so that effects of protostar formation may not be easily measurable by 
polarization maps.




\acknowledgments
We thank the referee for constructive comments that aided the discussion.
B.E. acknowledges support from a Natural Sciences and Engineering
Research Council of Canada (NSERC) Undergraduate Summer Research Award. S.B.
acknowledges support from an NSERC Discovery Grant. 

\bibliographystyle{apj}
\bibliography{myrefs}

\end{document}